# Molecular Signature of Polyoxometalates in Electron Transport of Silicon-based Molecular Junctions

Maxime Laurans,[a] Kevin Dalla Francesca,[a] Florence Volatron,*[a] Guillaume Izzet,[a] David Guerin,[b] Dominique Vuillaume,*[b] Stéphane Lenfant *[b] and Anna Proust*[a]

*a) Sorbonne Université, CNRS, Institut Parisien de Chimie Moléculaire, IPCM, 4 Place Jussieu, F-75005 Paris, France,*
*b) Institute for Electronics Microelectronics and Nanotechnology (IEMN), CNRS, Univ. Lille, Av. Poincaré, Villeneuve d'Ascq, France.*

Corresponding authors : *anna.proust@sorbonne-universite.fr*, *stephane.lenfant@iemn.univ-lille1.fr*, *dominique.vuillaume@iemn.fr*

Polyoxometalates (POMs) are unconventional electro-active molecules with a great potential for applications in molecular memories, providing efficient processing steps onto electrodes are available. The synthesis of the organic-inorganic polyoxometalate hybrids $[PM_{11}O_{39}\{Sn(C_6H_4)C\equiv C(C_6H_4)N_2\}]^{3-}$ (M = Mo, W) endowed with a remote diazonium function is reported together with their covalent immobilization onto hydrogenated n- Si(100) substrates. Electron transport measurements through the resulting densely-packed monolayers contacted with a mercury drop as a top electrode confirms their homogeneity. Adjustment of the current-voltage curves with the Simmon's equation gives a mean tunnel energy barrier $\Phi_{POM}$ of 1.8 eV and 1.6 eV, for the Silicon-Molecules-Metal (SMM) junctions based on the polyoxotungstates (M = W) and polyoxomolybdates (M =Mo), respectively. This follows the trend observed in the electrochemical properties of POMs in solution, the polyoxomolybdates being easier to reduce than the polyoxotungstates, in agreement with lowest unoccupied molecular orbitals (LUMOs) of lower energy. The molecular signature of the POMs is thus clearly identifiable in the solid-state electrical properties and the unmatched diversity of POM molecular and electronic structures should offer a great modularity.

## Introduction

Polyoxometalates (POMs) are nano-scaled transition metal oxides displaying discrete and reversible multi-reduction processes with minimal structural rearrangement.[1-5] This makes them promising candidates to be integrated into multi-level non-volatile molecular memories.[6,7,8,9] The transport properties of self-assembled monolayers or mutilayer films of $H_3[PW_{12}O_{40}]$ electrostatically deposited onto 3-aminopropyl-triethoxysilane (APTES)-modified $SiO_2$/Si surfaces have been characterized in lateral or vertical molecular junctions.[6,10,11] Recently the Dawson type $[W_{18}O_{54}(SeO_3)_2]^{4-}$ POM was incorporated into a flash memory transistor by drop casting around the Si nanowire channel covered with $SiO_2$.[12,13] Surprisingly, theoretical calculations concluded that the intimate nature of the POMs is not so important since devices integrating the parent $[W_{18}O_{56}(WO_6)_2]^{10-}$ were expected to behave similarly, despite the increase of the total charge of the polyanion. In the reported examples the ultimate electrical properties depend upon the number of POM layers or the POM packing density. To improve the performances a better control of the POM/electrode interface is required. This includes controlling the POM packing density and the homogeneity of the monolayer, the POM ordering and the POM deposition mode onto the electrode (electrostatic versus covalent anchoring).

    Ideally, a weak electronic coupling between the molecules and electrodes is required to favor a long data retention time of the memory (low electron transfer rate). In case of weak molecule/electrode coupling, the electronic structure of the molecules in the solid-state molecular junction is not strongly perturbed compared to the one known in solution and the electron transport properties in the junctions is mainly governed by the properties of the molecules.[14,15] A relationship can then be established between the energetics of the molecules in the solid-sate devices and in solution.[16,17] Similarly, the asymmetries in voltage-dependent electron transport properties (*i.e.* more efficient for one of two voltage polarities) can be related to the spatial and energy position of the molecular orbital involved in the electron transport.[18-20] By contrast, a strong molecule/electrode coupling results in severe modification of the electronic structure (molecule orbital energy shift and broadening),[21] up to the extreme case of electrode Fermi energy pinning,[22,23] for which the electron transport in the molecular junction is no longer dependent on the nature of the molecules.[24]

    Owing to their great structural diversity POMs are well-suited to address the issue of the relationship between transport properties and chemical and electronic molecular structures. In this contribution, we have thus chosen two Keggin-type POMs differing only in the nature of their constitutive metals, molybdenum

versus tungsten. Both have been functionalized by the same organic tether with a remote diazonium function for covalent anchoring onto silicon substrates, yielding isostructural and isocharged hybrids. In solution molybdates are known to be reduced more easily than their tungstates analogues, which corresponds to lowest unoccupied molecular orbitals (LUMOs) of lower energy. To which extend is this trend transposable and identifiable in solid-state molecular junctions? Can we control the energetics of the interface and the charge transport by substituting Mo for W? To study these issues, we describe the covalent grafting of two POM hybrids TBA$_3$[PM$_{11}$O$_{39}$\{Sn(C$_6$H$_4$)C≡C(C$_6$H$_4$)N$_2$\}] ( M = W, named **K$^W_{Sn}$[N$_2^+$]** and M = Mo, named **K$^{Mo}_{Sn}$[N$_2^+$]**) onto hydrogenated n-Si(100). Integration of monolayers of electroactive molecules in electronic memory devices has been a topic of increasing interest [25] [26] [27,28] and studies that have made use of oxide-free silicon as bottom electrode are particularly relevant to the CMOS approach. [29] [30-32] [33] [34] Different procedures have been considered to immobilize molecules onto the hydrogenated Si-substrate obtained after etching of the native oxide, among which hydrosilylation,[35-37] and, to a lesser extent, spontaneous grafting from aromatic diazonium.[38-40] Dediazonation results in the formation of robust Si-C covalent bonds and densely packed monolayers of POMs connected to the substrate by conjugated tethers. These POM monolayers onto Si have been contacted by a Hg drop to form Silicon-Molecules-Metal (SMM) junctions and the current-voltage (I-V) curves have been recorded, from which we have deduced the electronic structure (molecular energy level) of the SMM junctions. We have found that the trend on the extracted LUMO energy level in the solid-state SMM is following the trend of the POM reduction potentials as determined electrochemically in solution.

## Results and discussion
### Synthesis of TBA$_3$[PW$_{11}$O$_{39}$\{Sn(C$_6$H$_4$)C≡C(C$_6$H$_4$)N$_2$\}] (K$^W_{Sn}$[N$_2^+$])

Applications of POM-based materials in molecular nanosciences are still limited by their processing steps. Indeed, POMs are polyanions with counter cations, so that most examples of POM deposition onto electrodes rely on cation exchange[11,41-45] or entrapment into polymeric matrices.[46,47] Materials prepared by drop-casting or spin-coating have also been reported.[48-50] Another strategy is to prepare organic-inorganic POM hybrids with a remote anchor able to react either with complementary functions on a pre-assembled SAM (two-step approach) or directly onto the electrode.[51] [52] [53] The two-step approach has been exemplified by the formation of metal-alkoxide bonds or the more common peptide bonds.[54,55] [56,57] Some of us have investigated the direct grafting and have reported on the grafting of POM hybrids terminated with protected thiols or disulfide onto Au.[58,59] Among the possible anchors the diazonium function has especially retained our attention because it (i) allows grafting onto various substrates, Au, carbonaceous materials and Si [21,60] and (ii) because dediazonation results in the formation of strong covalent bonds with the substrates, resistant to the application of wide range of external bias.[61-63] Yet formation of the diazonium function will introduce an aromatic ring in the organic tether. In previous work, we have thus described the grafting of the diazonium functionalized polyoxotungstate TBA$_3$[PW$_{11}$O$_{39}$\{Ge(C$_6$H$_4$)C≡C(C$_6$H$_4$)N$_2$\}] (**K$^W_{Ge}$[N$_2^+$]**, TBA stands for tetra-butylammonium cations) onto glassy carbon and hydrogenated silicon and studied the kinetics of electron transfer at the modified electrode in solution.[64] [65] As charge storage nodes, polyoxomolybdates offer more perspectives than their tungsten counterparts because the cost of their reduction processes is lower (higher redox potentials).[47,66-69] This prompted us to prepare TBA$_3$[PMo$_{11}$O$_{39}$\{Sn(C$_6$H$_4$)C≡C(C$_6$H$_4$)N$_2$\}] (**K$^{Mo}_{Sn}$[N$_2^+$]**), to study its electro-grafting onto glassy carbon and discuss the electrochemical behavior of the corresponding modified electrodes. However **K$^W_{Ge}$[N$_2^+$]** and **K$^{Mo}_{Sn}$[N$_2^+$]** are differing not only in the nature of the constitutive metal W/Mo but also in the nature of the metalloid substituent Ge/Sn. For synthetic reasons, we failed to prepare the exact analog **K$^{Mo}_{Ge}$[N$_2^+$]**. Although the redox properties of **K$^W_{Sn}$** derivatives are expected to be almost indistinguishable from those of **K$^W_{Ge}$** (as verified for example for TBA$_4$[PW$_{11}$O$_{39}$\{GeC$_6$H$_4$I\}] **K$^W_{Ge}$[I]** and TBA$_4$[PW$_{11}$O$_{39}$\{SnC$_6$H$_4$I\}] **K$^W_{Sn}$[I]**) direct comparison of the properties of modified electrodes obtained from **K$^W_{Ge}$[N$_2^+$]** on one hand, and **K$^{Mo}_{Sn}$[N$_2^+$]** on the other hand is not completely satisfactory. In this contribution we thus describe the synthesis of the missing member of the diazonium-terminated POM family TBA$_3$[PW$_{11}$O$_{39}$\{Sn(C$_6$H$_4$)C≡C(C$_6$H$_4$)N$_2$\}] (**K$^W_{Sn}$[N$_2^+$]**). As silicon is technologically relevant to the field of molecular electronics, in an hybrid molecular semi-conductor approach compatible with CMOS (Complementary metal oxide semi-conductor) technology,[70] [71] [72] **K$^W_{Sn}$[N$_2^+$]** and **K$^{Mo}_{Sn}$[N$_2^+$]** have been covalently immobilized onto n-Si(100). Probing the charge transport across the POM layer onto silicon by contacting it with a metal electrode to form a SMM junction is the next step on the road to nanoelectronic devices involving POMs. Furthermore, comparison of the electrical behavior of the W/Mo POM junctions will contribute to the understanding of the POM role in the charge transport process.

The iodo-aryl terminated hybrids **K^M[I]** (M = W, Mo) are valuable platforms providing a large range of available POM building blocks for subsequent integration in complex architectures and assemblies.[73] Synthetic paths exploit the common Sonogashira cross-coupling reaction that has been adapted to the POM hybrids chemistry.[73-76] As depicted on Scheme 1, TBA$_4$[PW$_{11}$O$_{39}${Sn(C$_6$H$_4$)C≡C(C$_6$H$_4$)N$_3$Et$_2$}] (**K^W_{Sn}[N_3Et_2]**) was first prepared in 55 % yield by reaction between **K^W_{Sn}[I]** and an excess of 3,3-diethyl-1-(4-ethynylphenyl)triaz-1-ene acting as a protected diazonium group in DMF and subsequent work-up. It was characterized by IR, $^1$H and $^{31}$P NMR, ESI-MS and elemental analysis (see experimental part and ESI). The second step consisted in the deprotection of the triazene group to release the diazonium function, achieved by adding an excess of trifluoroactic acid (TFA) to a solution of **K^W_{Sn}[N_3Et_2]** in acetonitrile.

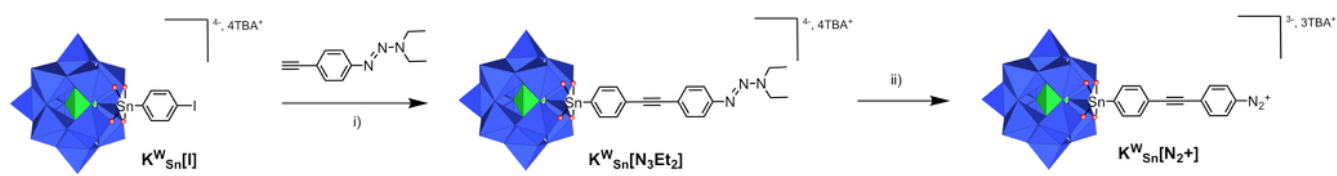

**Scheme 1.** *Synthetic route to the POM hybrid **K^W_{Sn}[N_2^+]**. In this polyhedral representation, the WO$_6$ octahedra are depicted with oxygen atoms at the vertices and metal cations buried inside. Color code: WO$_6$ octahedra, blue; PO$_4$ tetrahedra, green. i) [Pd(PPh$_3$)$_2$Cl$_2$], CuI, Et$_3$N, DMF overnight ii) TFA 10 min in MeCN.*

Addition of diethylether was carried out in the presence of tetrabutylammonium hexafluorophosphate to recover the desired product as a tetrabutylammonium salt and limit its protonation. The diazonium-terminated hybrid **K^W_{Sn}[N_2^+]** was thus obtained as a yellow powder in 79% yield and characterized by IR, $^{31}$P and $^1$H NMR, with integration of the signals of the $^1$H NMR spectrum consistent with 8 protons of the aromatic tether for 3 TBA cations. Unfortunately, no reliable ESI-MS spectrum or elemental analysis could be obtained because of the instability of the diazonium function. For this reason it is also recommended to use a freshly prepared sample. Besides the characteristic fingerprint of the [PW$_{11}$O$_{39}${SnAr}]$^{4-}$ core, the IR spectrum of **K^W_{Sn}[N_2^+]** displays a weak band at 2255 cm$^{-1}$ that we assigned to the N≡N stretch of the diazonium function (a weak band at 2208 cm$^{-1}$ was attributed to the C≡C bond).

### Grafting of K^W_{Sn}[N_2^+] and K^{Mo}_{Sn}[N_2^+] onto hydrogenated n-Si(100)

The just-etched Si-H substrate was thus dip-coated into a fresh solution of the diazonium-terminated POM **K^W_{Sn}[N_2^+]** or **K^{Mo}_{Sn}[N_2^+]**, without any supplied external bias. We have previously shown that the immersion time and rinsing procedure were critical to get a uniform monolayer with a high packing density and to avoid the formation of multilayer by competitive electrostatic deposition of extra POMs.[65] Because the diazonium route yields strong Si-C bonds, thorough rinsing and sonication is feasible without degradation of the POM monolayer, which would not be possible in the case of an electrostatic deposition of the POMs. This is an an important asset of the covalent immobilization of POMs. Formation of multilayers by radical attack on the already bound aromatics is a limitation of diazonium route, albeit strategies to control the layer growth have been devised, based on bulky substituents or the presence of radical scavengers.[77, 78] Steric hindrance is probably provided here by the POM itself. Indeed, ellipsometry measurements confirm the formation of a monolayer with a mean thickness of 2.7 and 3.2 nm for the **K^W_{Sn}** and **K^{Mo}_{Sn}** derivatized monolayers respectively. This is in good agreement with a total thickness expected between 2.5 and 3.4 nm given that the diameter of the POM is ~1 nm, the length of the organic tether is around 1.4 nm and that the TBA cations should contribute for 0.5 nm to 1 nm, depending on the conformation of the butyl chains, and tacking into account the possible 30° bending of the layer. Ellipsometry measurements were reproducible on several samples and the homogeneity of the layer was further probed by AFM in tapping mode (Fig. 1), which gave smooth images with a root-mean-square (RMS) roughness of 0.15 nm and 0.19 nm for the **K^W_{Sn}** and **K^{Mo}_{Sn}** samples respectively, slightly superior to the n-Si substrate with a RMS roughness measured at ~ 0.10 nm and in good accordance with our previous work.[65]

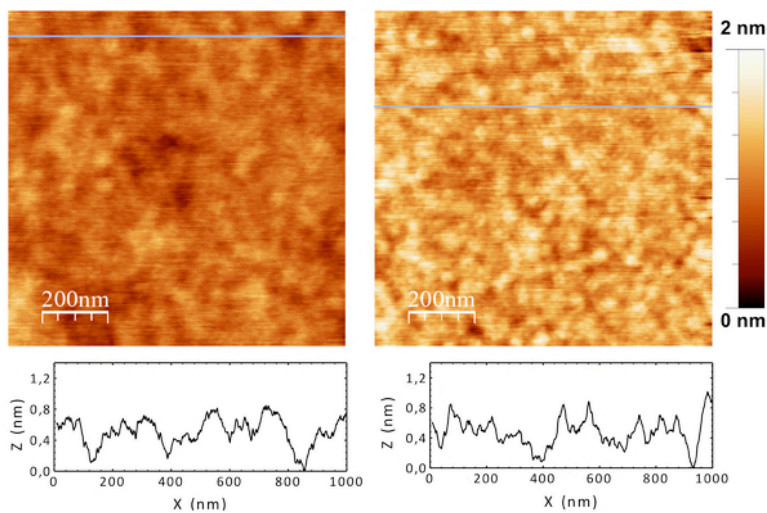

***Fig. 1*** *Tapping-mode AFM images (up) (the scale bar corresponds to 100 nm) and height profile following the line in the AFM image (bottom) of **$K^W_{Sn}$** (left) and **$K^{Mo}_{Sn}$** (right) modified n-Si(100) substrates.*

Both samples were also studied by XPS and all constitutive elements of the POMs were detected: Sn, P, C, N, O and the metals, W for $K^W_{Sn}$ and Mo for $K^{Mo}_{Sn}$ modified substrates. A part of the high resolution spectra are presented in Figure 2. For the $K^W_{Sn}$ compound, the spectrum for the 4f level of the element shows the typical spin orbit doublet at 36.1 ($W4f_{7/2}$) and 38.2 eV ($W4f_{5/2}$) corresponding to oxidized W(VI) atoms, as reported before.[56, 65] This is confirmed by the presence of oxygen atoms bonded to metallic species, as attested by the contribution at 531.0 eV on the O1s deconvoluted peak (see - S11). The $Sn3d_{5/2}$ peak at 487.1 eV is in good agreement with oxidized Sn(IV) species as previously reported on tin oxide solids or tin derivative POM hybrid films [79, 80] (the $Sn3d_{3/2}$ peak, not presented here for the sake of clarity, is usually spoiled by satellite peaks related to alkaline ions). The N1s and C1s photopeaks are quite similar to the ones of a diazonium terminated POM reference powder drop-casted on a silicon substrate (see Figure S12).

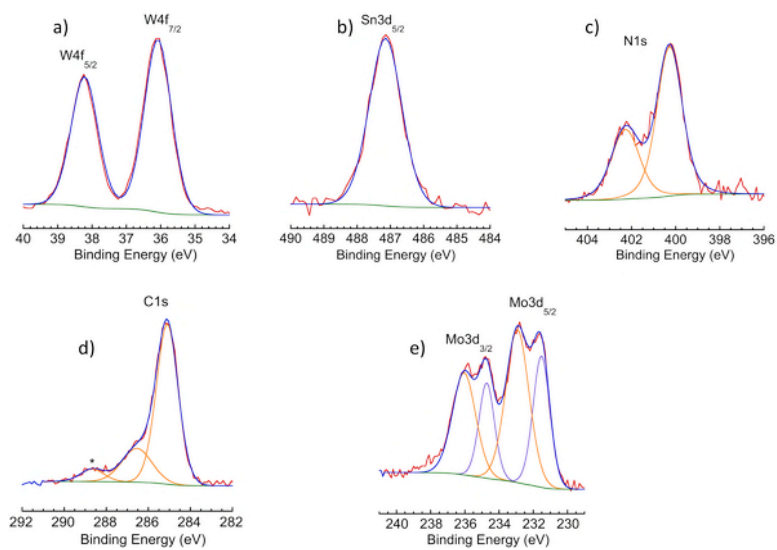

***Fig. 2.*** *High resolution X-ray photoelectron spectra for (a) W4f (b) $Sn3d_{5/2}$ (c) N1s (d) C1s in a **$K^W_{Sn}$** monolayer and for (e) Mo3d in a **$K^{Mo}_{Sn}$** monolayer. The peak marked with an asterisk corresponds to a contamination by carbon atoms in an oxidized state.*

On the N1s spectrum, one more contribution appears at 400.3 eV that is attributed to confined TBA counter-cations surrounding the grafted POMs. For the $K^{Mo}_{Sn}$ sample, the P2p, Sn3d, O1s, N1s and C1s spectra exhibit the same features as for the $K^W_{Sn}$ sample (see Figure S13). However, on the Mo3d spectrum, in addition to the expected Mo(VI) spin orbit doublet at 233.0 eV ($3d_{5/2}$) and 236.1 eV ($3d_{3/2}$), a new doublet at 231.5 ($3d_{5/2}$) and 234.7 eV ($3d_{3/2}$) was observed and attributed to Mo(V) atoms.[80] This means that around 40% of the grafted

POMs are reduced. To understand the origin of this reduction, we compare the Mo3d spectra of a same batch of a **K$^{Mo}_{Sn}$[N$_2^+$]** powder, drop-casted on a Si/SiO2 substrate, and used to covalently graft the **K$^{Mo}_{Sn}$** species on hydrogenated silicon (see Figure S14). We observe that the amount of Mo(V) is far lower (10%) in the powder reference than in the **K$^{Mo}_{Sn}$** layer. This means that the main contribution for the Mo reduction does come neither from the chemical synthesis, nor from the X-ray irradiation during XPS measurement. The Mo reduction may occur during the grafting on silicon. Indeed, the reduction potential of the polyoxomolybdate at -0.5 V vs SCE is close to the reduction potential of the diazonium group. Thus the hydrogenated silicon substrate, reducing agent for the diazonium function of the **K$^{Mo}_{Sn}$[N$_2^+$]** is also able to partially reduce the Mo(VI) atoms of the inorganic core.

**Current-voltage (I-V) curves**

Current-voltage (I-V) curves were measured by contacting the POM monolayer by a Hg drop acting as the top electrode in a glove box filled under a nitrogen flow. The drop was gently brought into contact with the sample surface thanks to a camera. The voltage V was applied on the Hg drop and the highly doped (degenerated, resistivity of 1-5 x 10$^{-3}$ Ω.cm) n-type ⟨100⟩ silicon substrate is grounded through the ammeter to measure the current (Fig. 3a).

Figures 3-b and 3-c show 2D histograms of the 75 I-V curves measured on the SMM junctions with the mercury drop technique for the **K$^{Mo}_{Sn}$** and **K$^{W}_{Sn}$** derivatized monolayers, respectively. The 75 I-V traces were acquired at different location on the POM monolayer surface, with 1 to 5 traces per location, and they were not averaged. The two curves are quite similar with a low dispersion, inferior to one decade reflecting the homogeneity of the monolayer grafting on the substrate. However, we notice a slight difference between these two curves. The I–V curves obtained on the **K$^{W}_{Sn}$** monolayer are symmetric with respect to the voltage polarity, while the **K$^{Mo}_{Sn}$** monolayer present an absolute current value slightly higher at +1V than at -1V. However, the asymmetry is weak (rectification ratio I$_{+1V}$/I$_{-1V}$ is around 2.5, see Figure S15) and is not significant.[19, 20]

We analyzed the I-V curves using the Simmons model to extract the tunnel energy barrier of the monolayer placed between two electrodes.[81-84, 85, 86]
For highly doped Si, there is no Schottky barrier related to the semi-conductor and only tunneling through the molecular monolayer is operating, with a tunneling barrier height Φ$_{POM}$.

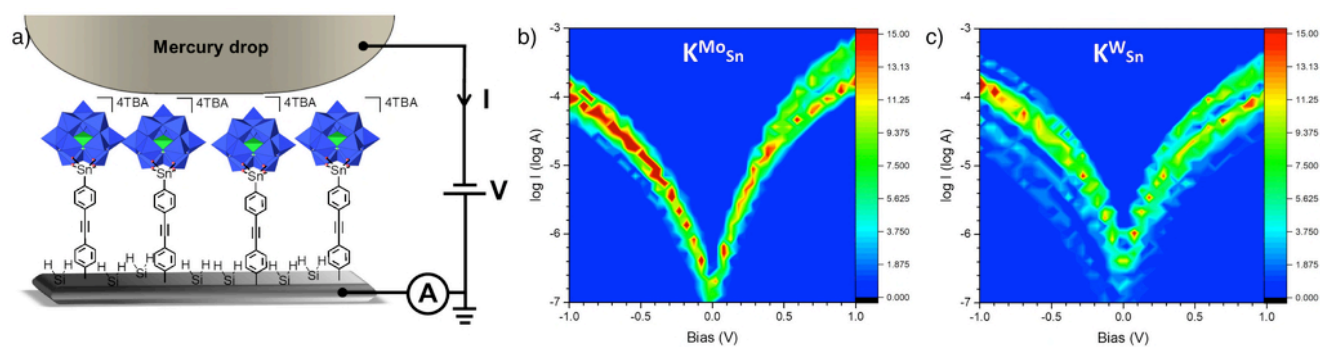

*Fig. 3. a): Scheme of the molecular layer grafted on a Si substrate and the SMM junction Si-K$^{W}_{Sn}$//Hg. (b) 2D current histogram of 75 I–V curves measured for K$^{Mo}_{Sn}$ and (c) K$^{W}_{Sn}$ monolayers chemically grafted on a highly doped (degenerated) n-type ⟨100⟩ silicon substrates (resistivity of ∼1-5 x 10$^{-3}$ Ω.cm). Voltages were applied on the mercury and Si substrate was grounded.*

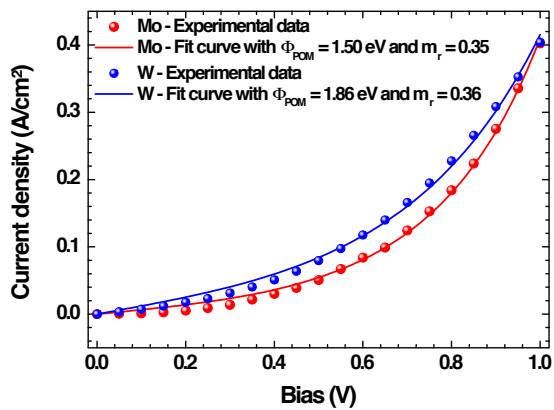

***Fig. 4.*** *Typical I-V adjustment between 0 and 1 V with the Simmon's equation for the $K^{Mo}_{Sn}$ (red curve) and $K^{W}_{Sn}$ (blue curve) derivatized monolayers. Adjustments of the Simmons equation with the experimental data for positive bias are good (R > 0.98). The other known parameters are : thickness 2.7 nm and 3.2 nm (W and Mo derivatives, respectively), contact area $3 \times 10^{-4}$ cm².*

We extracted the barrier height and the effective mass of the monolayer directly by adjusting the I-V curve with the Simmon's equation[82] (See Supporting Information for more details on this adjustment) since the other parameters are known: the thickness of the tunnel barrier corresponds to the monolayer thickness measured by ellipsometry and the surface contact area is estimated to ~ $3 \times 10^{-4}$ cm² for the mercury drop technique used (see details in ESI).

A representative I-V curve and the corresponding adjustment with the Simmon's equation are presented on Figure 4 for the **$K^{Mo}_{Sn}$** and **$K^{W}_{Sn}$** derivatized monolayers respectively. Histograms of the $\Phi_{POM}$ values extracted from these adjustments for the whole set of 75 I-V curves are shown on Figure 5a (histograms of the fitted effective mass are given in ESI, Figure S16). The histograms of $\Phi_{POM}$ are fitted by a Gaussian law, and there is a clear offset towards higher values for the **$K^{W}_{Sn}$** monolayers compared to the **$K^{Mo}_{Sn}$** ones, with mean values of $\Phi_{POM}$ = 1.8 eV (standard deviation 0.26 eV) and $\Phi_{POM}$ = 1.6 eV (standard deviation 0.35 eV), respectively. We note that these distributions include the raw current data distribution (Fig. S15 in ESI) and the uncertainty of the measured thicknesses and contact area. From these average barrier height values for the both monolayers, we deduced the energy diagram of the junction (Fig. 5b), where the LUMO is localized at $\Phi_{POM}$ above the silicon Fermi energy (at 4.2 eV below the vacuum level). This value is an average between the LUMO of the POM unit itself and the one of the π-conjugated tether (double energy barrier, see ESI). However, since the tether is the same, the observed shift of $\Phi_{POM}$ is ascribed to the different chemical nature of the POM. These average values $\Phi_{POM}$ for both monolayers are, in the following, compared with the redox potentials determined by cyclic-voltammetry.

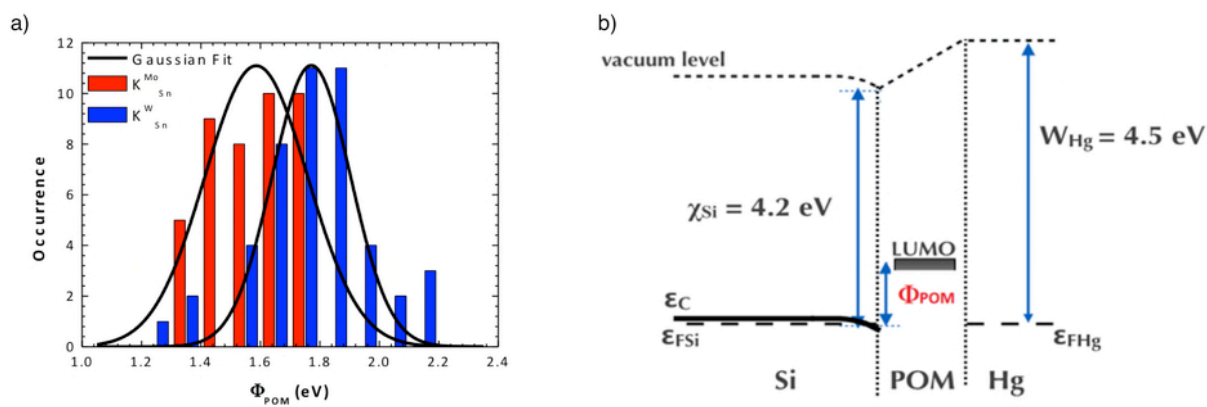

***Fig. 5.*** *(a) Barrier height ($\Phi_{POM}$) histograms obtained by Simmon's adjustments on the different I-V curves, and adjusted by a Gaussian distribution for the $K^{Mo}_{Sn}$ and $K^{W}_{Sn}$ derivatized monolayers. (b) Schematic energy diagram of the Si-POM//Hg drop as deduced from the I–V measurements.*

**Effect of the molecular structure of the POM**

Significantly different mean values of the energy barrier $\Phi_{POM}$ have been inferred from the I-V measurements for the $K^{W}_{Sn}$ derivatized layers compared to the $K^{Mo}_{Sn}$ ones, which indicates that the POMs are indeed involved in the charge transport mechanism. Furthermore, in solution the reduction processes associated to the polyoxomolybdates are shifted to higher potentials compared to those of the analogous polyoxotungstates.[87-89] The cyclic-voltammetry of $K^{Mo}_{Sn}[I]$ versus $K^{W}_{Sn}[I]$ and $K^{Mo}_{Sn}[N_3Et_2]$ versus $K^{W}_{Sn}[N_3Et_2]$ is presented in ESI as an illustration, together with the cyclic-voltammetry of the $K^{W}_{Sn}$-derivatized layer and the relevant data gathered in Table S1. They show that (i) the redox mid-point potentials $E_{1/2}$ of molybdates are indeed about 500 mV higher than those of the corresponding tungstates, at about -0.5 V/SCE compared to -1.0 V/SCE; (ii) the nature of the tether has almost no influence on the $E_{1/2}$ values as we have already noticed on other POM hybrids;[90-92] (iii) the $E_{1/2}$ values for the POMs confined at the electrode are similar to those determined for the POMs in solution.[58, 64, 65] An empirical linear correlation between the φ and $E_{1/2}$ values has been proposed in the literature.[20, 93] However, at variance with ferrocene, for example, the E potentials of POMs are dependent upon several parameters such as the solvent[94] and the counter-cations,[95, 96] which might introduce some deviation. On a qualitative point of view, the matching between the relative $\Phi_{POM}$ values of $K^{W}_{Sn}$ and $K^{Mo}_{Sn}$ and their redox potential scale is however satisfying, the higher $\Phi_{POM}$ the more negative $E_{1/2}$. Note that in the case of the $K^{Mo}_{Sn}$ SMM, the value of $\Phi_{POM}$ is an average between the two oxidation states, Mo(V) and Mo(VI) observed by XPS.

# Conclusions

Two diazonium-terminated POM hybrids $TBA_3[PW_{11}O_{39}\{Sn(C_6H_4)C\equiv C(C_6H_4)N_2\}]$ ($K^{W}_{Sn}[N_2^+]$) and $TBA_3[PMo_{11}O_{39}\{Sn(C_6H_4)C\equiv C(C_6H_4)N_2\}]$ ($K^{Mo}_{Sn}[N_2^+]$) differing only by the nature of their constitutive metals (W/Mo) have been prepared and covalently immobilized onto hydrogenated n-Si(100) by dediazonation to prepare $K^{W}_{Sn}$- and $K^{Mo}_{Sn}$- modified electrodes respectively. Solid-state molecular junctions have been closed by a Hg-top electrode and I-V curves have been recorded for the resulting SMM junctions. The I-V curves have been analyzed using the Simmons model to extract the tunneling barrier heights $\Phi_{POM}$, which correspond to the position of the LUMOs relative to the silicon Fermi energy. The clear difference between the mean values of $\Phi_{POM}$ for the $K^{W}_{Sn}$- and $K^{Mo}_{Sn}$- modified electrodes, 1.8 eV and 1.6 eV respectively, follows the trend of the POM reduction potentials E as determined electrochemically in solution, with a higher energy barrier $\Phi_{POM}$ related to lower (more negative) redox potential E. We were thus able to detect the molecular signature of the POMs in the SMM junctions and this gives some insights into the energetics of the interface. Indeed, these results demonstrate a weak electronic coupling between the POMs and the Si substrate, making these electro-active molecules prone for molecular memory applications.

# Experimental section

Chemicals and solvents were supplied from Aldrich or Acros and used as received, except triethylamine and acetonitrile that were distilled from $CaH_2$. $TBA_4[PW_{11}O_{39}\{SnC_6H_4I\}]$ ($K^{W}_{Sn}[I]$)[97] and $TBA_3[PMo_{11}O_{39}\{Sn(C_6H_4)C\equiv C(C_6H_4)N_2\}]$ ($K^{Mo}_{Sn}[N_2^+]$)[98] were prepared as previously reported (TBA stands for the tetrabutylammonium cation).

NMR spectra were recorded on a Bruker AvanceIII Nanobay 400 MHz spectrometer equipped with a BBFO probehead. $^1H$ chemical shifts are quoted as parts per million (ppm) relative to tetramethylsilane using the solvent signals as secondary standard (s: singlet, d: doublet, t: triplet, sex: sextet, m: multiplet) and coupling constants (J) are quoted in Hertz (Hz). $^{31}P$ chemical shifts are quoted relative to 85% $H_3PO_4$. IR spectrum of the powder was recorded from a KBr pellet on a Jasco FT/IR 4100 spectrometer. High-resolution ESI mass spectra were recorded using an LTQ Orbitrap hybrid mass spectrometer (Thermofisher Scientific, Bremen, Germany) equipped with an external ESI source operated in the negative ion mode. Spray conditions included a spray voltage of 3 kV, a capillary temperature maintained at 280 °C, a capillary voltage of −30 V, and a tube lens offset of −90 V. Sample solutions in acetonitrile (10 pmol.µL$^{-1}$) were infused into the ESI source by using a syringe pump at a flow rate of 180 µL.h$^{-1}$. Mass spectra were acquired in the Orbitrap analyzer with a theoretical mass resolving power ($R_P$) of 100 000 at m/z 400, after ion accumulation to a target value of 10$^5$ and a m/z range detection from m/z 300 to 2000. All data were acquired using external calibration with a mixture of caffeine, MRFA peptide and Ultramark 1600 dissolved in Milli-Q water/ HPLC grade acetonitrile (50/50, v/v). Elemental analyses were performed at the Institut de Chimie des Substances Naturelles, Gif sur Yvette, France.

**Synthesis of $TBA_4[PW_{11}O_{39}\{Sn(C_6H_4)C\equiv C(C_6H_4)N_3(C_2H_5)_2\}]$ $K^{W}_{Sn}[N_3Et_2]$**

$K^{W}_{Sn}[I]$ (199 mg, 0.050 mmol), 3,3-diethyl-1-(4-ethynylphenyl)triaz-1-ene ((33.5 mg, 0.166 mmol)), bis-(triphenylphosphine)palladium(II) dichloride (5.4 mg, 0.008 mmol) and copper iodide (1.4 mg, 0.007 mmol) were dissolved in dried and purged DMF (5 mL). The mixture was purged 5 min and freshly distilled triethylamine (150 µL, 1.133 mmol) was added. After stirring 24 hrs at room temperature, diethylether (40 mL) was added to precipitate the desired product. The recovered precipitate was dissolved in a minimum of TBABr solution (173.1 mg, 0.54 mmol in 10 mL acetonitrile). Finally, the pure product was obtained after precipitation by absolute ethanol (30 mL) and diethyl ether (2x30 mL) as beige powder (114.2 mg 55 %).

$^1$H NMR (400 MHz, CD$_3$CN): δ(ppm) 7,71 (d, $^3J_{H-H}$=8,12Hz, $^3J_{Sn-H}$= 96,42Hz, 2H, Ar-*H*), 7,61 (d, $^3J_{H-H}$=8,12Hz, $^4J_{Sn-H}$= 32,72Hz, 2H, Ar-*H*), 7,52 (d, $^3J_{H-H}$=8,68Hz, 2H, Ar-*H*), 7,38 (d, $^3J_{H-H}$=8,68Hz, 2H, Ar-*H*), 3,79 (q, $^3J_{H-H}$=7,12Hz, 4H, N-C*H*$_2$-CH$_3$), 3,12 (m, 32H, N-C*H*$_2$-CH$_2$-CH$_2$-CH$_3$), 1,63 (m, 32H, N-CH$_2$-C*H*$_2$-CH$_2$-CH$_3$), 1,39 (sex, $^3J_{H-H}$=7,36Hz, 32H, N-CH$_2$-CH$_2$-C*H*$_2$-CH$_3$), 1,26 (m, 6H, N-CH$_2$-C*H*$_3$), 0,98 (t, $^3J_{H-H}$=7,36Hz, 48H, N-CH$_2$-CH$_2$-CH$_2$-C*H*$_3$); $^{31}$P NMR (121 MHz, CD$_3$CN): δ (ppm) - 10,97 (s+d, $^2J_{Sn-P}$=23,49Hz); IR (KBr pellet, cm$^{-1}$): ν = 2962 (m), 2931 (m), 2866 (m), 1482 (m), 1379 (w), 1334 (w), 1239 (w), 1071 (m), 968 (s), 884 (m), 806 (vs), 512 (w), 384 (m). HRMS (ESI-): m/z: [M]$^{4-}$ calcd 768.07 found 768.07 ; [M+TBA]$^{3-}$ calcd 1104.86 found 1104.86; [M+2TBA]$^{2-}$ calcd 1778.48 found 1778.93 for C$_{18}$H$_{19}$SnN$_3$O$_{39}$PW$_{11}$. Anal. calcd for PSnW$_{11}$O$_{39}$C$_{82}$H$_{162}$N$_7$ (%): C 24.36, H 4.04, N 2.44; found : C 24.05, H 3.98, N 2.33.

**Synthesis of TBA$_3$[PW$_{11}$O$_{39}${Sn(C$_6$H$_4$)C≡C(C$_6$H$_4$)N$_2$}] K$^W_{Sn}$[N$_2^+$]**

Trifluoroacetic acid (TFA, 3.8µL, 0.050 mmol) was slowly added to a solution of **K$^W_{Sn}$[N$_3$Et$_2$]** (40.9 mg, 0.010 mmol) in dried acetonitrile (2 mL) and the solution was stirred at room temperature for 10 minutes. A brown unidentified solid was filtered off the yellow solution and and tetrabutylammonium hexafluorophosphate (TBAPF$_6$, 77.5 mg, 0.200 mmol) was added to the filtrate. The desired product was precipitated by dropping the filtrate into diethylether (20 mL). The yellow powder was dried at air for 30 minutes (31.1 mg, 79%). **K$^W_{Sn}$[N$_2^+$]** is stored in a sealed flask at -30°C.

$^1$H NMR (400 MHz, CD$_3$CN, 298°K): δ(ppm): 8.51 (d, $^3J_{H-H}$=9.10Hz, 2H, Ar-*H*), 7.93 (d, $^3J_{H-H}$=9.10Hz, 2H, Ar-*H*), 7.75 (d, $^3J_{H-H}$=8.23Hz, 2H, Ar-*H*), 7.71 (d, $^3J_{H-H}$=8.23, 2H, Ar-*H*), 3.13 (m, 24H, N-C*H*$_2$-CH$_2$-CH$_2$-CH$_3$), 1.62 (m, 24H, N-CH$_2$-C*H*$_2$-CH$_2$-CH$_3$), 1.37 (sex, $^3J_{H-H}$=7.33Hz, 24H, N-CH$_2$-CH$_2$-C*H*$_2$-CH$_3$), 0.99 (t, $^3J_{H-H}$=7.33Hz, 36H, N-CH$_2$-CH$_2$-CH$_2$-C*H*$_3$); $^{31}$P NMR (121,5 MHz, CD$_3$CN, 298°K): δ(ppm) -10.96 (s+d, $^2J_{Sn-P}$=23,90Hz); IR (KBr pellet, cm$^{-1}$): ν = 2960 (m), 2933 (m), 2872 (m), 2255 (w), 2208 (w), 1571 (m), 1482 (m), 1380 (w), 1070 (m), 963 (s), 886 (m), 795 (vs), 515 (w), 380 (m), 332 (w).

**Surface covalent grafting**

Pieces of silicon wafer (Highly phosphorous-doped n-Si(100) wafers purchased from Siltronix resistivity 1-5x10$^{-3}$ Ω.cm) were sonicated for 5 min in dichloromethane and rinsed with absolute ethanol. They were then dipped two times in baths with the following compositions (i) 15 min piranha solution (mixture 2:1 of concentrated sulphuric acid and 30 wt% hydrogen peroxide solution) and (ii) 2 minutes in 5 % hydrofluoric acid (HF). After the last bath of HF, the hydrophobic dried substrate was directly dipped into a fresh 10$^{-3}$ M solution of **K$^M_{Sn}$[N$_2^+$] (M = W or Mo)** in distilled and degased acetonitrile. The system was left under inert atmosphere at room temperature for about 1 hour (the dipping time was adjusted depending on the ellipsometry results), upon which the substrate was thoroughly rinsed with a flux of acetonitrile and sonicated for 1 minute in a bath of acetonitrile. Finally the substrate was dried under nitrogen flow and preserved from oxidation under inert atmosphere.

*Caution: Piranha solution can be explosive in presence of organic compounds and HF is extremely toxic and corrosive. Manipulate with appropriate care.*

**Ellipsometry**

Ellipsometry measurements were performed on a 1*1 cm$^2$ sample and obtained using a monowavelength ellipsometer SENTECH SE 400 equipped with a He-Ne laser at λ = 632.8 nm. The incident angle was 70°. The values ns = 3.875 and ks = 0.018 were taken for the silicon wafer,[99] and ns = 1.48 and ks = 0 for the layer of POMs.[100] At least 6 measurements were performed on a same sample in different zones, to check the homogeneity of the layer. A mean value for the thickness was calculated when the standard deviation was lower than 0.2 nm.

**AFM**

The surface morphology of the POM monolayer was determined by imaging with a Dimension Icon atomic force microscope (AFM) from Bruker in tapping mode. Silicon cantilevers (Arrow™ NC from Nanoworld; k = 42 N.m$^{-1}$, radius < 10 nm) were used to acquire AFM images of 1 x 1 µm$^2$ at 1 Hz with a resolution of 512 x 512 pixels.

**XPS**

XPS analyses were performed using an Omicron Argus X-ray photoelectron spectrometer. The monochromated AlK$_α$ radiation source (*h*ν = 1486.6 eV) had a 300 W electron beam power. The emission of photoelectrons from the sample was analyzed at a takeoff angle of 90° under ultra-high vacuum conditions (≤ 10$^{-10}$ Torr). Spectra were carried out with a 100 eV pass energy for the survey scan and 20 eV pass energy for the C1s, O1s, N1s regions. Binding energies were calibrated against the Si2p binding energy at 99.4 eV and element peak intensities were corrected by Scofield factors. The spectra were fitted using Casa XPS v.2.3.15 software (Casa Software Ltd., U.K.) and applying a Gaussian/Lorentzian ratio G/L equal to 70/30.

**I-V measurements**

We used a hanging mercury drop as top electrode, to contact electrically the POM monolayer grafted on a highly doped n-Si substrate. Calibrated mercury drops (99.9999%, purchased from Fluka) were generated by a controlled growth mercury electrode system (CGME model from BASi) placed inside a glove box purged with a nitrogen flow. The mechanical contact between the sample and the hanging mercury drop was formed by moving up a precision lab-lift (supporting the sample) under the control of a digital video camera. The electrical contact area estimated by capacitance measurement on calibrated samples (Si/SiO$_2$ structure with a precisely known SiO$_2$ thickness of 10 nm and permittivity of 3.9 as measured by spectroscopic ellipsometry) is around $3 \times 10^{-4}$ cm$^2$. The current–voltage measurements were done with an Agilent (Keysight) 4156C parameter analyzer. The voltage was always applied on the Hg top electrode, and the current measured at the grounded Si substrate.

# Acknowledgements

KDF thanks Sorbonne Universities and the program PERSUE for financial support.

# Electronic Supporting Information

Molecular Signature of Polyoxometalates in Electron Transport of Silicon-based Molecular Junctions

*Maxime Laurans, Kevin Dalla Francesca, Florence Volatron,\* Guillaume Izzet, David Guerin, Dominique Vuillaume,\* Stéphane Lenfant,\* Anna Proust\**

**Content**

1. NMR and ESI-MS characterization of TBA$_4$[PW$_{11}$O$_{39}${Sn(C$_6$H$_4$)C≡C(C$_6$H$_4$)N$_3$(C$_2$H$_5$)$_2$}] **K$^W_{Sn}$[N$_3$Et$_2$]** and TBA$_3$[PW$_{11}$O$_{39}${Sn(C$_6$H$_4$)C≡C(C$_6$H$_4$)N$_2$}] **K$^W_{Sn}$[N$_2^+$]**

Figure S1. $^1$H (300.13 MHz) and $^{31}$P (121.5 MHz, framed inset) NMR spectra of **K$^W_{Sn}$[N$_3$Et$_2$]** in CD$_3$CN.

Figure S2. Comparison of experimental (upper trace) and calculated (lower trace) isotopic peaks for the most abundant ions

Figure S3. $^1$H (400.13 MHz) and $^{31}$P (121.5 MHz, framed inset) NMR spectra of K$^W_{Sn}$[N$_2^+$] in CD$_3$CN.

**2. Cyclic voltammetry**

Table S1. Mid-point redox potentials (versus SCE) from cyclic-voltammograms recorded at 100 mV s$^{-1}$ for POM hybrids in solution or covalently immobilized onto the working electrode.

Figure S4. Cyclic voltammogram of **K$^W_{Sn}$[N$_3$Et$_2$]** (1mM) at a GC electrode in a 0.1M TBAPF$_6$ solution in acetonitrile at scan rate of ν = 100 mV.s$^{-1}$.

Figure S5. Cyclic voltammogram of a **K$^W_{Sn}$**-functionalized glassy carbon electrode in a 0.1M TBAPF$_6$ solution in acetonitrile at scan rate of ν = 100 mV.s$^{-1}$.

Figure S6. Cyclic voltammograms of a **K$^W_{Sn}$**-functionalized glassy carbon electrode in a 0.1M TBAPF$_6$ solution in acetonitrile at scan rates of ν = 0.5, 1, 2, 3, 4, 5 V.s$^{-1}$

Figure S7. Cyclic voltammograms of a **K$^W_{Sn}$**-functionalized glassy carbon electrode in a 0.1M TBAPF$_6$ solution in acetonitrile at scan rate of ν = 0.4 V.s$^{-1}$

Figure S8. Cyclic voltammogram of a **K$^W_{Sn}$**-functionalized silicon electrode in a 0.1M TBAPF$_6$ solution in acetonitrile at scan rate of ν = 100 mV.s$^{-1}$.

Figure S9. Cyclic voltammogram of **K$^{Mo}_{Sn}$[N$_3$Et$_2$]** (1mM) at a GC electrode in a 0.1M TBAPF$_6$ solution in acetonitrile at scan rate of ν = 100 mV.s$^{-1}$.

Figure S10. Cyclic voltammogram of a $K^{Mo}_{Sn}$-functionalized glassy carbon electrode in a 0.1M TBAPF$_6$ solution in acetonitrile at scan rate of $v = 100$ mV.s$^{-1}$.

## 3. XPS Characterization

Figure S11. P2p and O1s HR-XPS spectra of $K^{W}_{Sn}$ modified n-Si(100) substrate

Figure S12. Sn3d$_{5/2}$, N1s, C1s and P2p HR-XPS spectra of a $K^{W}_{Mo}[N_2^+]$ reference powder drop-casted on a Si/SiO$_2$ substrate.

Figure S13. Sn3d$_{5/2}$, N1s, C1s, P2p and O1s HR-XPS spectra of $K^{Mo}_{Sn}$ modified n-Si(100) substrate.

Figure S14. Mo3d HR-XPS spectrum of (left) the $K^{Mo}_{Sn}[N_2^+]$ powder reference and (right) the $K^{Mo}_{Sn}$ modified n-Si(100) substrate.

## 4. Current histograms at -1V and 1V on $K^{W}_{Sn}$ and $K^{Mo}_{Sn}$ monolayers grafted on highly doped Si substrate

Figure. S15. Histograms of the measured current at +1V and -1V for the $K^{W}_{Sn}$ and $K^{Mo}_{Sn}$ samples.

## 5. I-V curves adjustment with the Simmon's equation

## 6. Histograms of effective mass obtained by Simmon's adjustments on $K^{W}_{Sn}$ and $K^{Mo}_{Sn}$ monolayers

Figure. S16. Reduced mass ($m_r$) histograms obtained by Simmon's adjustments on the different I-V curves, and adjusted by a Gaussian distribution for the $K^{Mo}_{Sn}$ and $K^{W}_{Sn}$ derivatized monolayers.

# 1. NMR and ESI-MS characterization of TBA$_4$[PW$_{11}$O$_{39}${Sn(C$_6$H$_4$)C≡C(C$_6$H$_4$)N$_3$(C$_2$H$_5$)$_2$}] K$^W_{Sn}$[N$_3$Et$_2$] and TBA$_3$[PW$_{11}$O$_{39}${Sn(C$_6$H$_4$)C≡C(C$_6$H$_4$)N$_2$}] K$^W_{Sn}$[N$_2^+$]

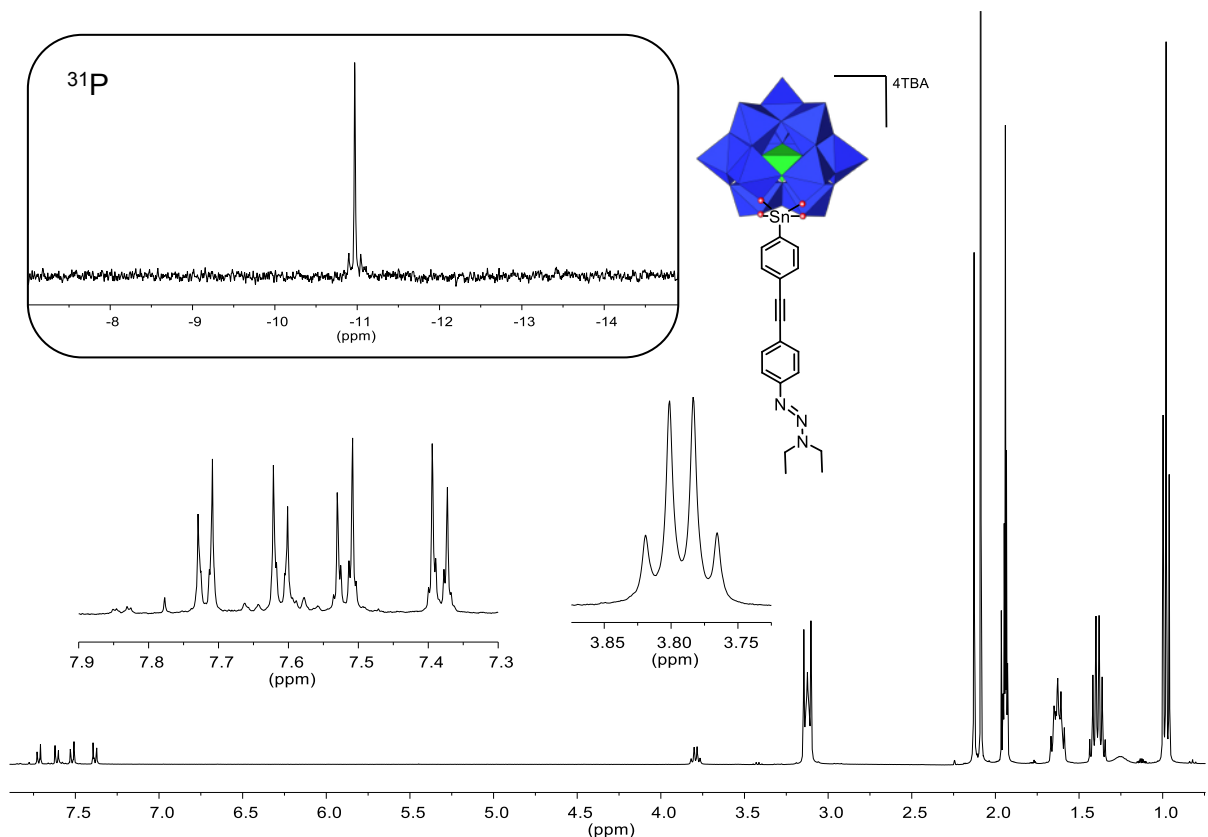

Figure S1. $^1$H (400 MHz) and $^{31}$P (121.5 MHz, framed inset) NMR spectra of **K$^W_{Sn}$[N$_3$Et$_2$]** in CD$_3$CN.

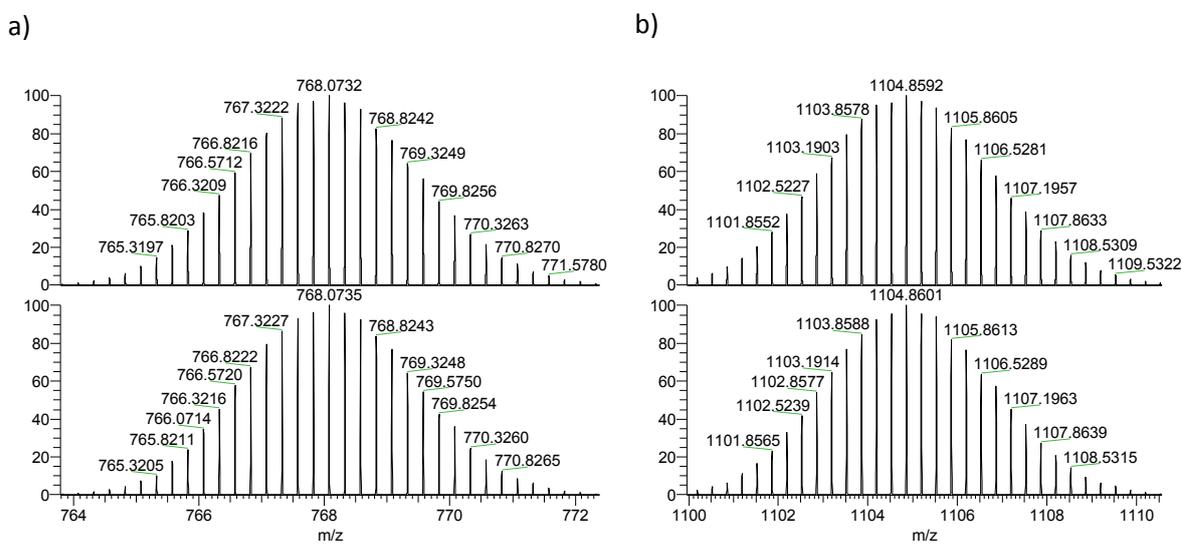

Figure S2. Comparison of experimental (lower trace) and calculated (upper trace) isotopic

peaks for the most abundant ions i.e. : a) [POM]$^{4-}$ calcd 768.07 found 768.07 ; b) [POM+TBA]$^{3-}$ calcd 1104.86 found 1104.86. POM = [PW$_{11}$O$_{39}${Sn(C$_6$H$_4$)C≡C(C$_6$H$_4$)N$_3$(C$_2$H$_5$)$_2$}]

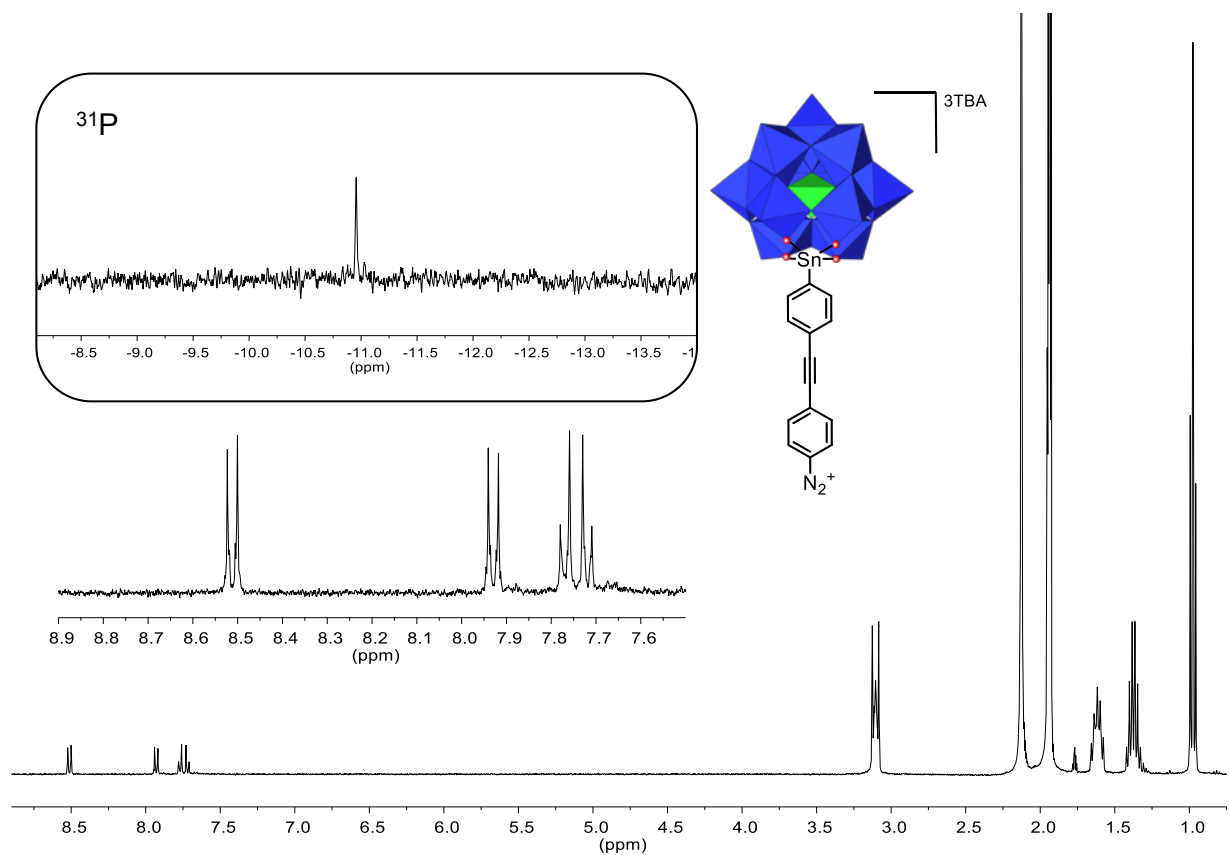

Figure S3. $^1$H (400 MHz) and $^{31}$P (121.5 MHz, framed inset) NMR spectra of **K$^W{}_{Sn}$[N$_2{}^+$]** in CD$_3$CN.

## 2. Cyclic voltammetry

**Cyclic voltammetry**

Electrochemical studies were performed on an Autolab PGSTAT 100 work station (Metrohm) using a standard 3-electrode setup filled with a 0.1 M solution of tetrabutylammonium hexafluorophosphate electrolyte in acetonitrile kept under argon. The electrochemical properties of the POMs in solution (1 mM) were investigated with a glassy carbon electrode (3 mm diameter) that was polished with 6 μm diamond paste, sonicated in ethanol for 5 min and dried with an argon flow. Alternatively the POM-modified silicon wafer itself was used as the working electrode. Platinum wire and saturated calomel electrode (SCE) equipped with a double junction were used as auxiliary and reference electrodes respectively. Grafting of $TBA_3[PM_{11}O_{39}\{Sn(C_6H_4)C\equiv C(C_6H_4)N_2\}]$ ($K^M_{Sn}[N_2^+]$) onto glassy carbon was achieved as previously described by cycling around the reduction wave of the diazonium function (between -0.2 and -0.8 V/SCE),[1,2] the modified electrode was then thoroughly rinsed and sonicated in DMF and acetonitrile before its characterization.

| V/SCE | $E_{p,red}$ | $E_{p,ox}$ | $E_{1/2}$= ½($E_{p,red}$+$E_{p,ox}$) |
|---|---|---|---|
| $TBA_4[PW_{11}O_{39}\{SnC_6H_4I\}]$ ($K^W_{Sn}[I]$) | | | -0.97 <br> -1.42 [3] |
| $TBA_4[PW_{11}O_{39}\{Sn(C_6H_4)C\equiv C(C_6H_4)N_3Et_2\}]$ ($K^W_{Sn}[N_3Et_2]$) | -1.07 | -0.94 | -1.00 |
| $K^W_{Sn}$-modified glassy carbon electrode | -1.03 | -1.00 | -1.01 |
| $K^W_{Sn}$-modified Si(100) electrode | -1.03 | -0.93 | -0.98 |
| | | | |
| $TBA_4[PMo_{11}O_{39}\{SnC_6H_4I\}]$ ($K^{Mo}_{Sn}[I]$) | | | -0,50 <br> -0,92 [4] |
| $TBA_4[PMo_{11}O_{39}\{Sn(C_6H_4)C\equiv C(C_6H_4)N_3Et_2\}]$ ($K^{Mo}_{Sn}[N_3Et_2]$) | | | -0.50 |
| $K^{Mo}_{Sn}$-modified glassy carbon electrode | | | -0.55 [2] <br> -0.52 |

Table S1. Mid-point redox potentials (versus SCE) from cyclic-voltammograms recorded at 100 mV s$^{-1}$ for POM hybrids in solution or covalently immobilized onto the working electrode.

For reasons that we do not fully explained at the moment it was not possible to recover well defined waves from the $K^{Mo}_{Sn}$-POMs grafted onto silicon. This might arise from a too fast growing of an insulting $SiO_2$ layer, which is very difficult to avoid when working in solution or from the presence of traces of protons, introduced in the last step of the synthesis of $K^{Mo}_{Sn}[N_2^+]$. This is particularly acute for molybdates that are more sensitive to protonation than their corresponding tungstates. Deliberate addition of protons results first in a broadening of the electrochemical waves and it is only at high proton concentrations that well-resolved waves are recovered, usually at higher potentials. This has been discussed in a previous contribution related to the immobilization of $K^{Mo}_{Sn}[N_2^+]$ onto glassy carbon electrode.[2] We have indeed observed that the shape of the electrochemical waves of the $K^{Mo}_{Sn}$-modified glassy carbon electrode was altered compared to the case of a $K^W_{Sn}$-modified glassy carbon electrode.[1,2]

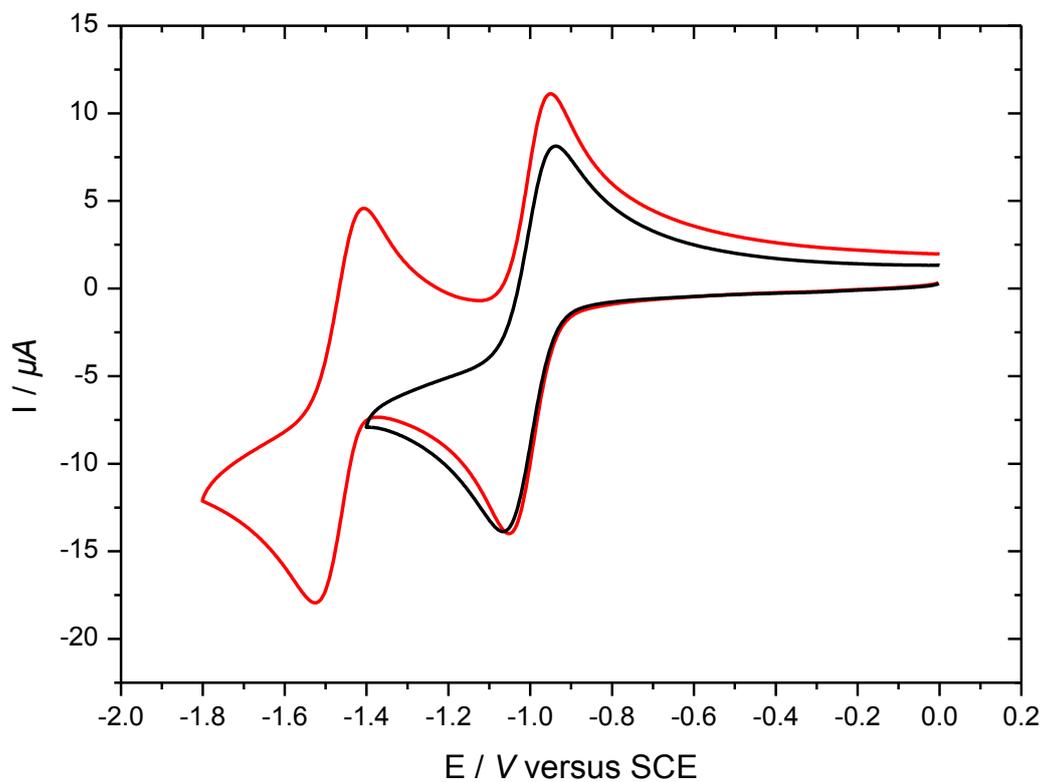

Figure S4. Cyclic voltammogram of $\mathbf{K^W_{Sn}[N_3Et_2]}$ (1mM) at a GC electrode in a 0.1M TBAPF$_6$ solution in acetonitrile at scan rate of $\nu$ = 100 mV.s$^{-1}$. $E_{1/2}^{red1}$ = -1.00V$_{/SCE}$

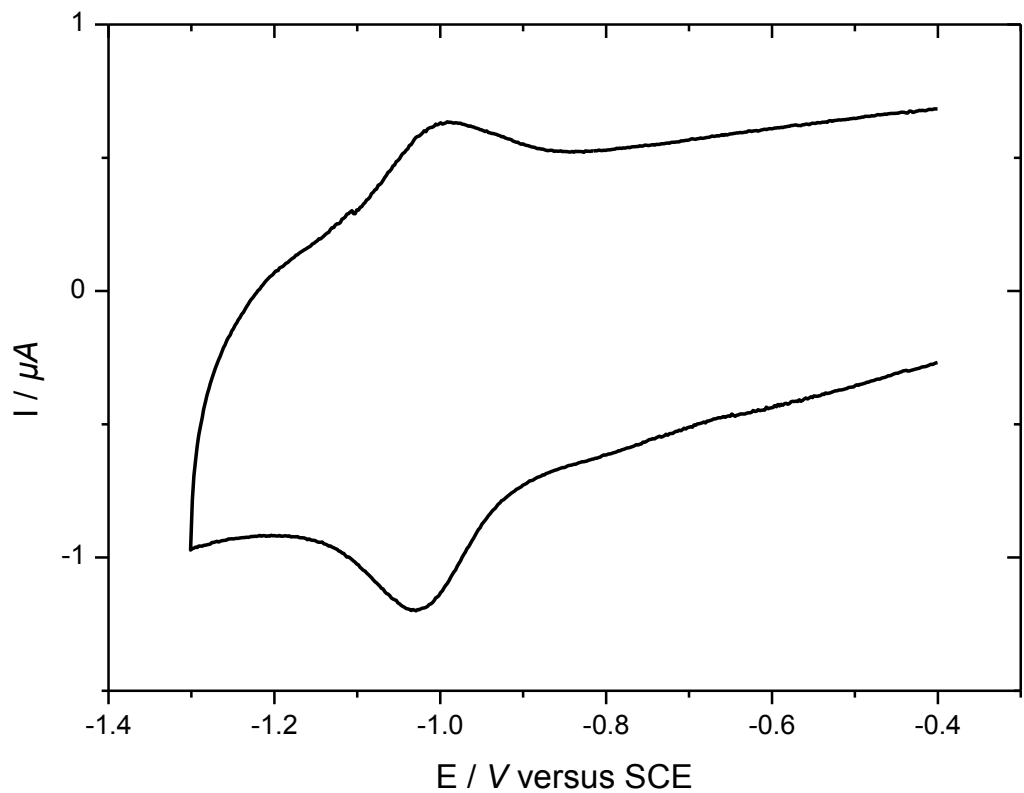

Figure S5. Cyclic voltammogram of a $K^W_{Sn}$-functionalized glassy carbon electrode in a 0.1M TBAPF$_6$ solution in acetonitrile at scan rate of $\nu$ = 100 mV.s$^{-1}$. $E_{1/2}^{red1}$ = -1.01V$_{/SCE}$

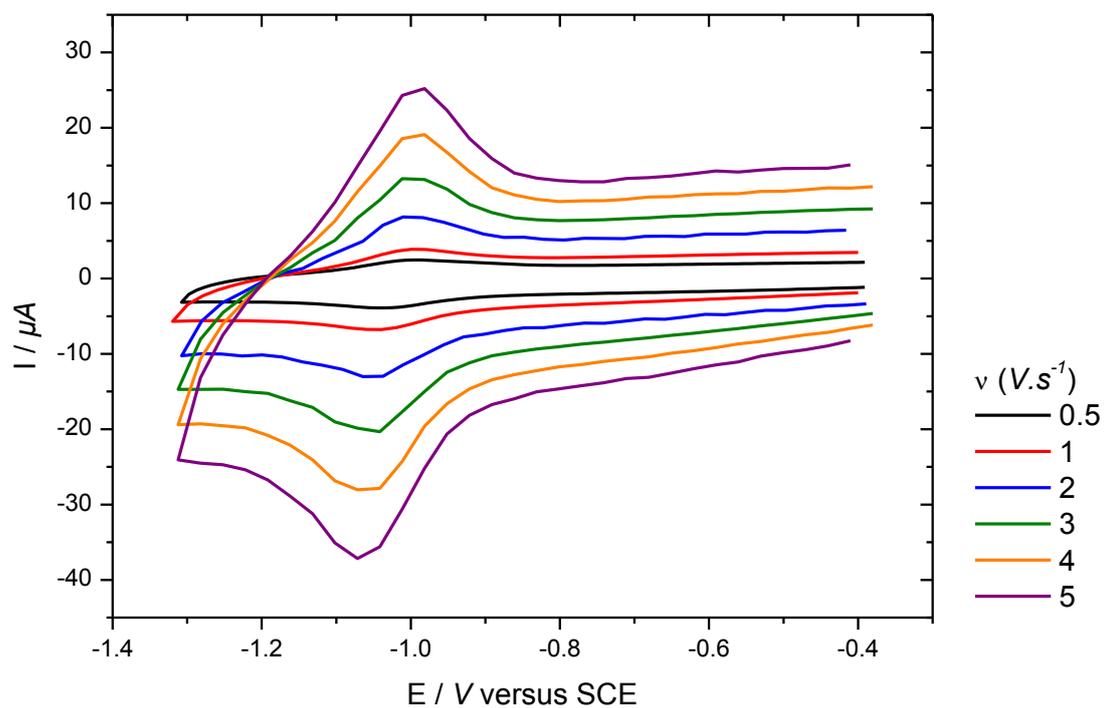

Figure S6. Cyclic voltammograms of a $K^{W}_{Sn}$-functionalized glassy carbon electrode in a 0.1M TBAPF$_6$ solution in acetonitrile at scan rates of ν = 0.5, 1, 2, 3, 4, 5 V.s$^{-1}$

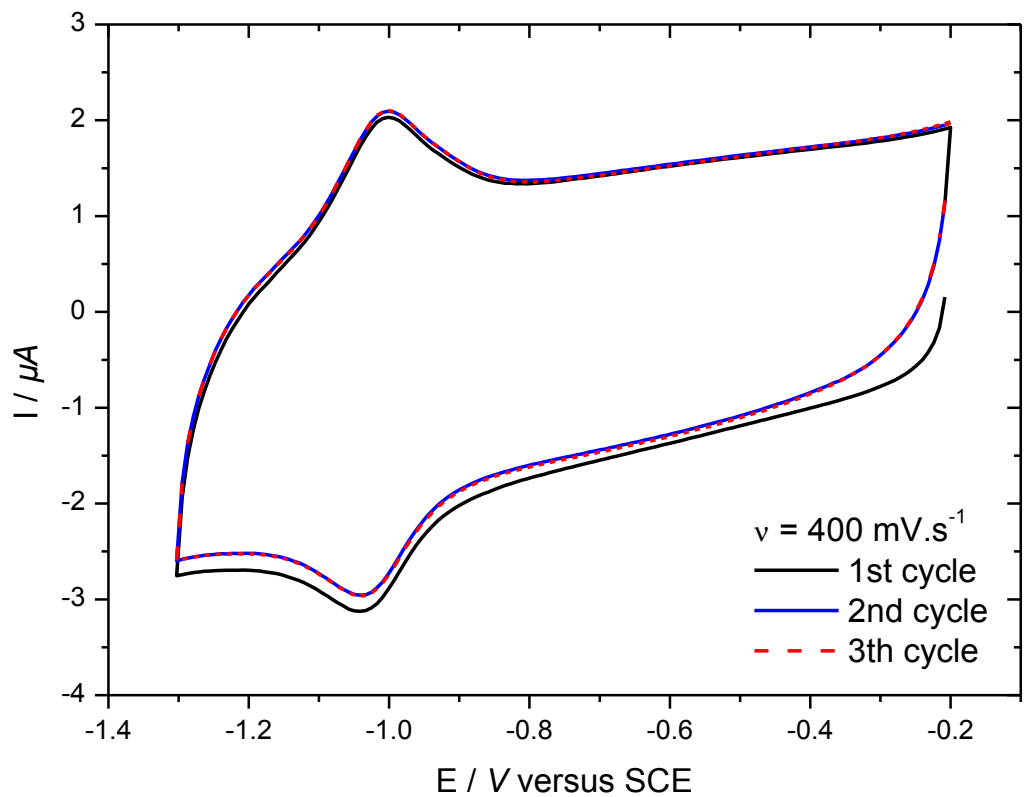

Figure S7. Cyclic voltammograms of a **K$^W_{Sn}$**-functionalized glassy carbon electrode in a 0.1M TBAPF$_6$ solution in acetonitrile at scan rate of ν = 0.4 V.s$^{-1}$

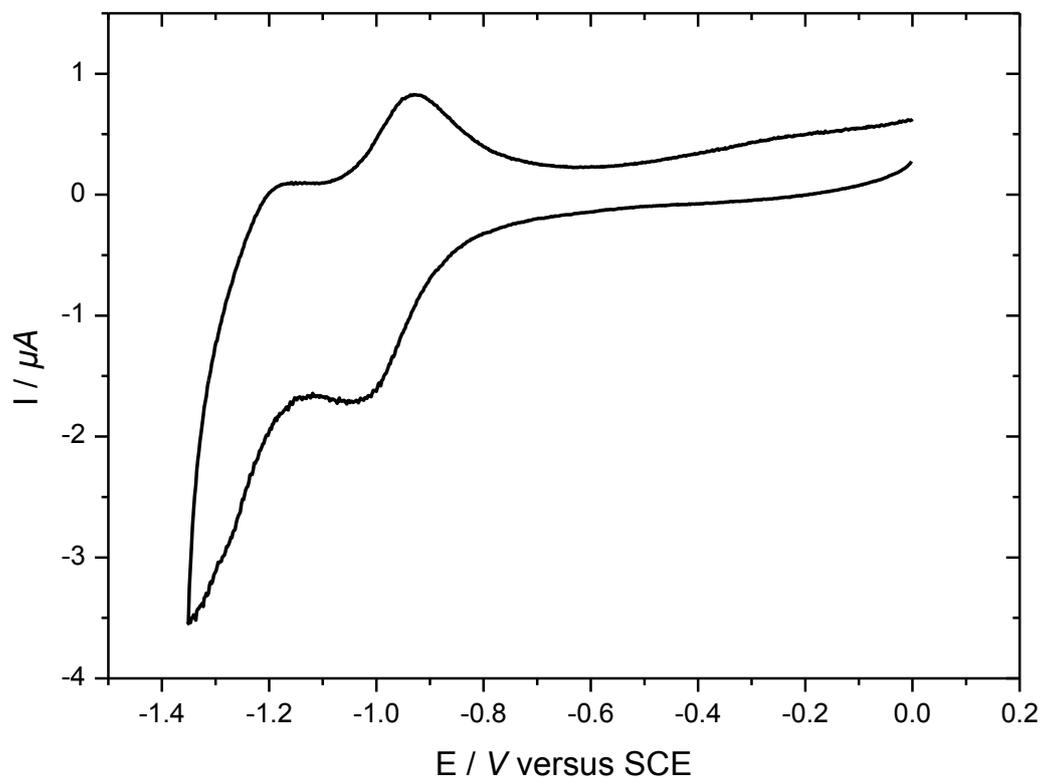

Figure S8. Cyclic voltammogram of a **K$^W_{Sn}$**-functionalized silicon electrode in a 0.1M TBAPF$_6$ solution in acetonitrile at scan rate of ν = 100 mV.s$^{-1}$. E$_{1/2}$ $^{red1}$ = -0.98 V$_{/SCE}$

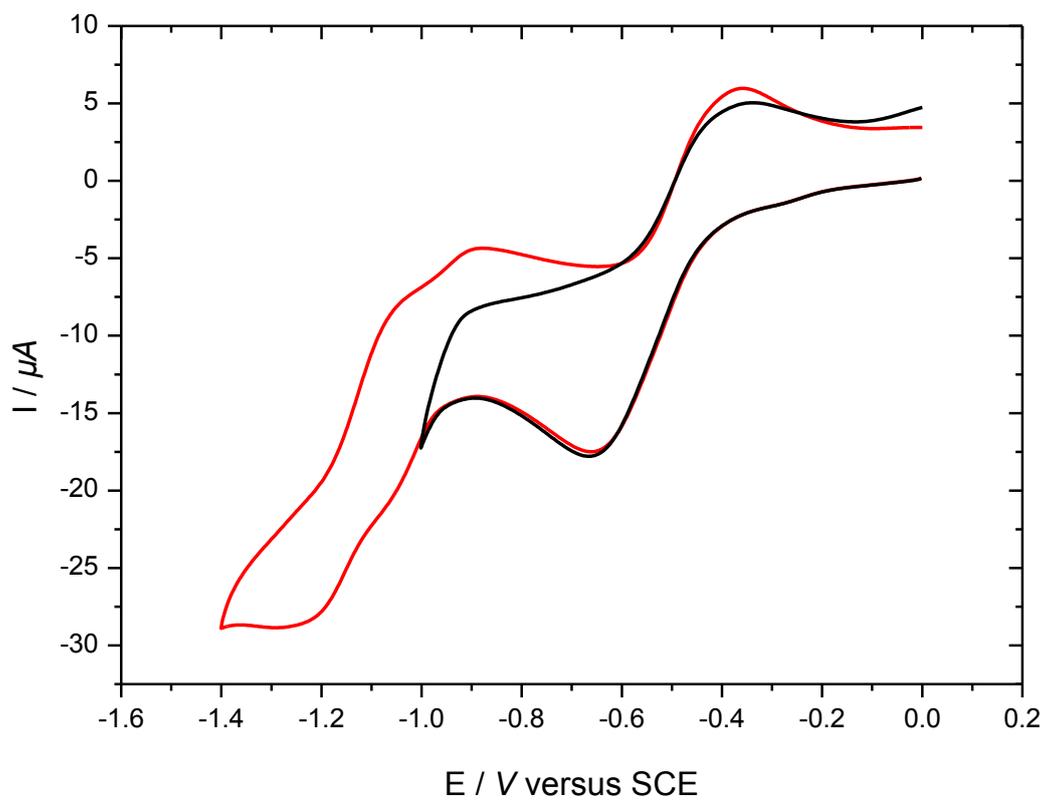

Figure S9. Cyclic voltammogram of **K$^{Mo}_{Sn}$[N$_3$Et$_2$]** (1mM) at a GC electrode in a 0.1M TBAPF$_6$ solution in acetonitrile at scan rate of ν = 100 mV.s$^{-1}$. E$_{1/2}$ $^{red1}$ = -0.50V$_{/SCE}$

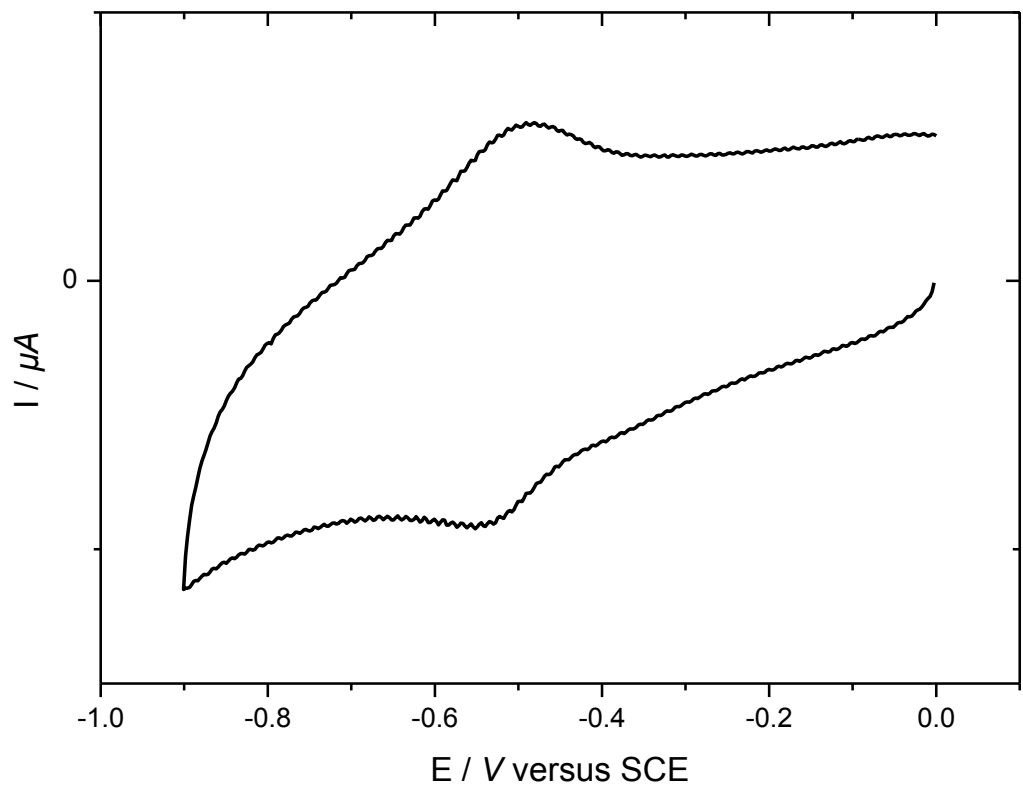

Figure S10. Cyclic voltammogram of a **K**<sup>**Mo**</sup><sub>**Sn**</sub>-functionalized glassy carbon electrode in a 0.1M TBAPF$_6$ solution in acetonitrile at scan rate of ν = 100 mV.s$^{-1}$. $E_{1/2}^{red1}$ = -0.52V$_{/SCE}$

## 3. XPS characterization

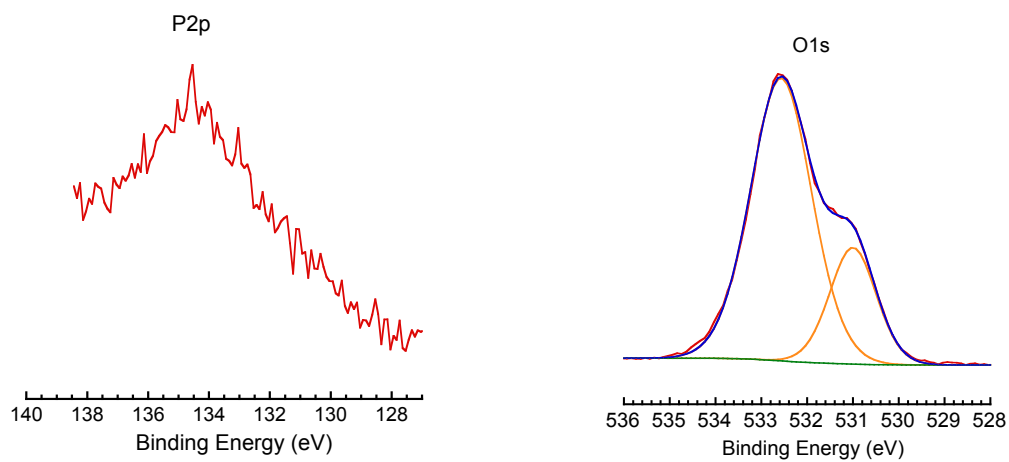

Figure S11. P2p and O1s HR-XPS spectra of $K^{W}_{Sn}$ modified n-Si(100) substrate. The contribution at 532.6 eV on the O1s photopeak is attributed to oxygen atoms in silicon dioxide. The XPS measurements were performed after numerous characterizations (included solid-state electrical measurements) and oxide may have been formed due to the repeated manipulation of the substrate.

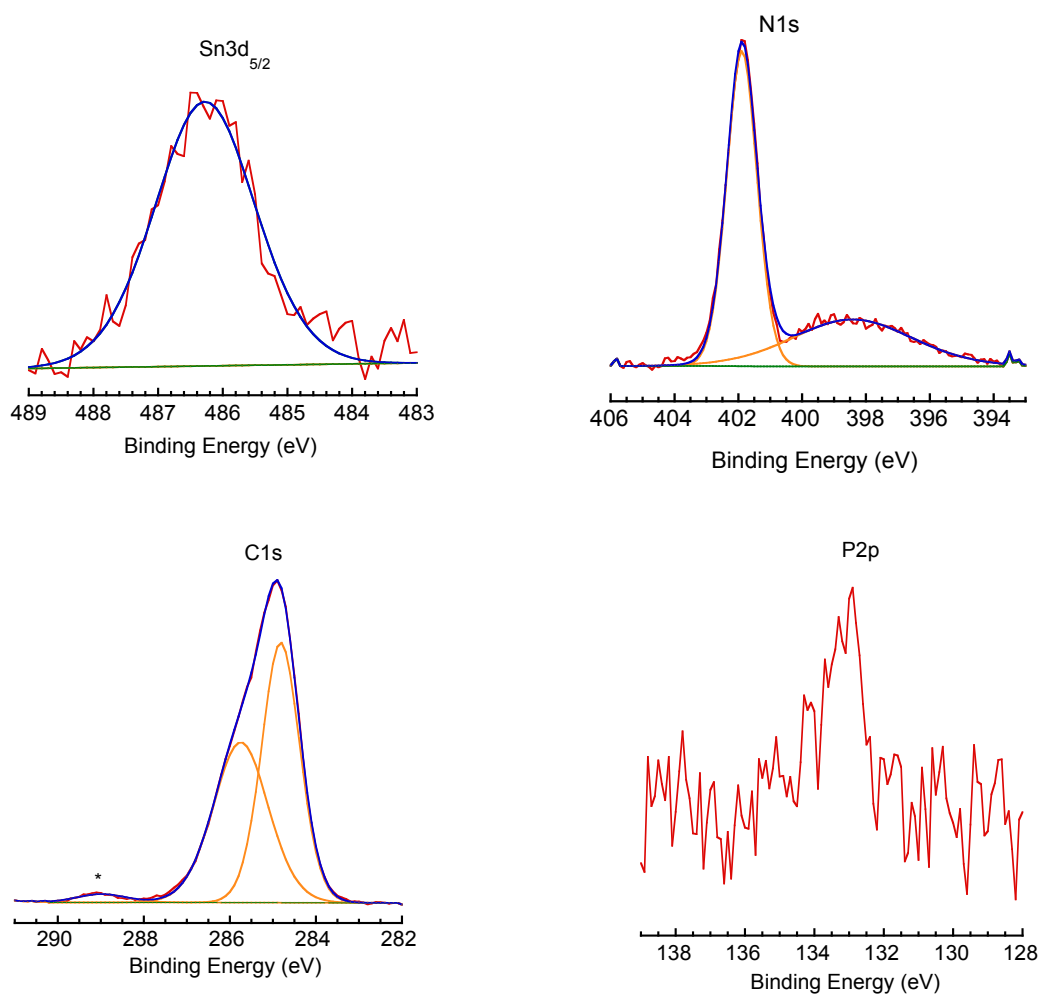

Figure S12. Sn3d$_{5/2}$, N1s, C1s and P2p HR-XPS spectra of a **K$^{Mo}_{Sn}$[N$_2^+$]** reference powder drop-casted on a Si/SiO$_2$ substrate. On the N1s spectrum, the broad peak at around 398 eV is attributed to nitrogen derivatives of the unstable diazonium group in the **K$^{Mo}_{Sn}$[N$_2^+$]** powder.

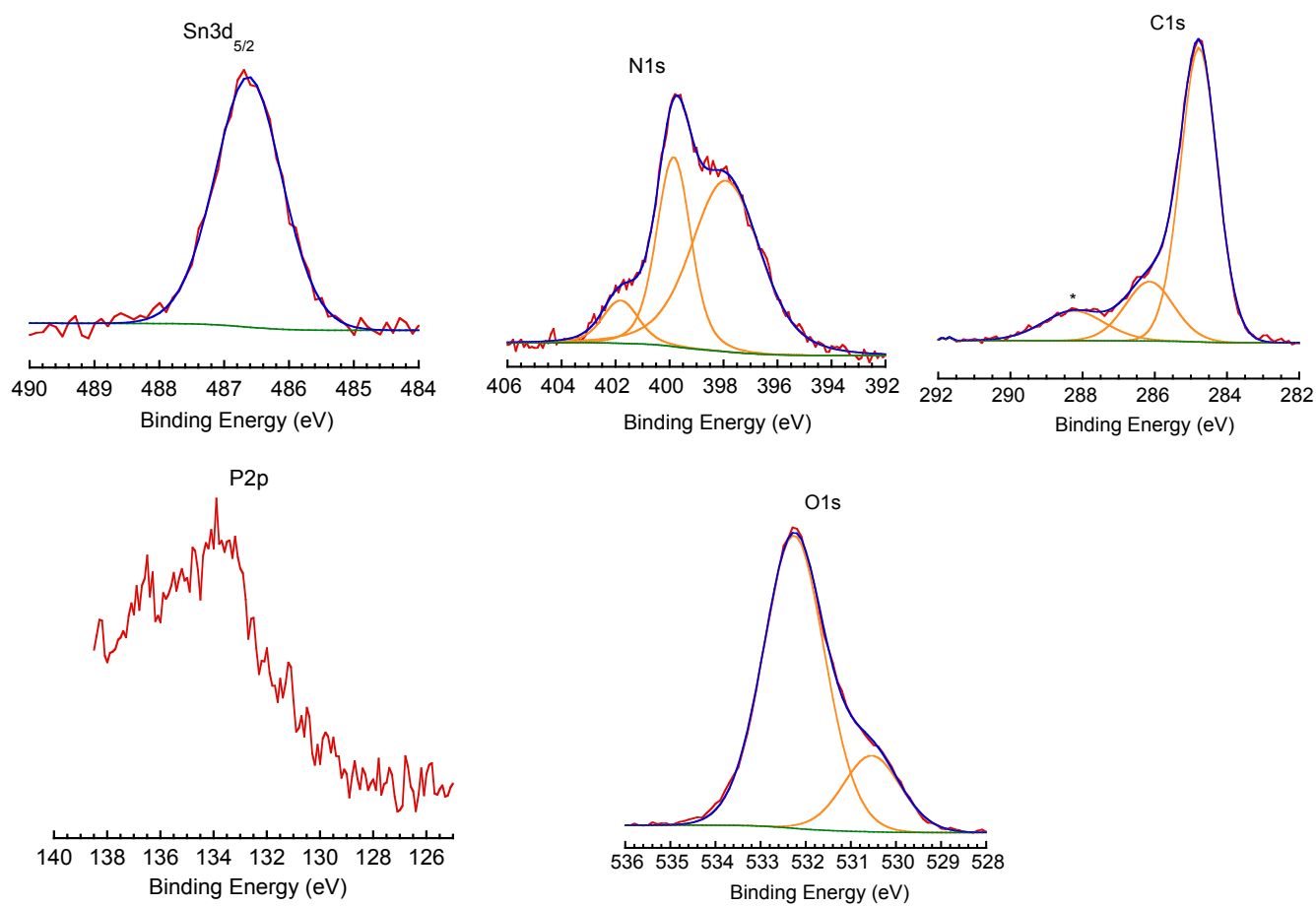

Figure S13. Sn3d$_{5/2}$, N1s, C1s, P2p and O1s HR-XPS spectra of **K$^{Mo}_{Sn}$** modified n-Si(100) substrate. The peak at 397.9 eV on the N1s spectrum corresponds to a satellite of Mo atoms.

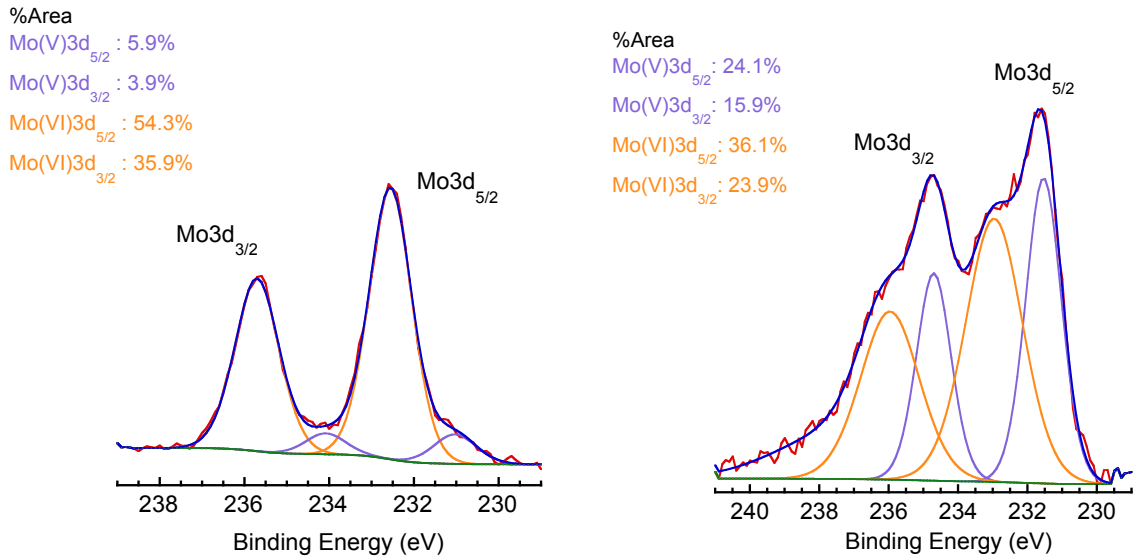

Figure S14. Mo3d HR-XPS spectra of (left) the $K^{Mo}_{Sn}[N_2^+]$ powder reference and (right) the $K^{Mo}_{Sn}$ modified n-Si(100) substrate. The same $K^{Mo}_{Sn}[N_2^+]$ batch was used to measure the powder reference and perform the grafting on hydrogenated silicon.

# 4. Current histograms at -1V and 1V on K$^W_{Sn}$ and K$^{Mo}_{Sn}$ monolayers grafted on highly doped Si substrate

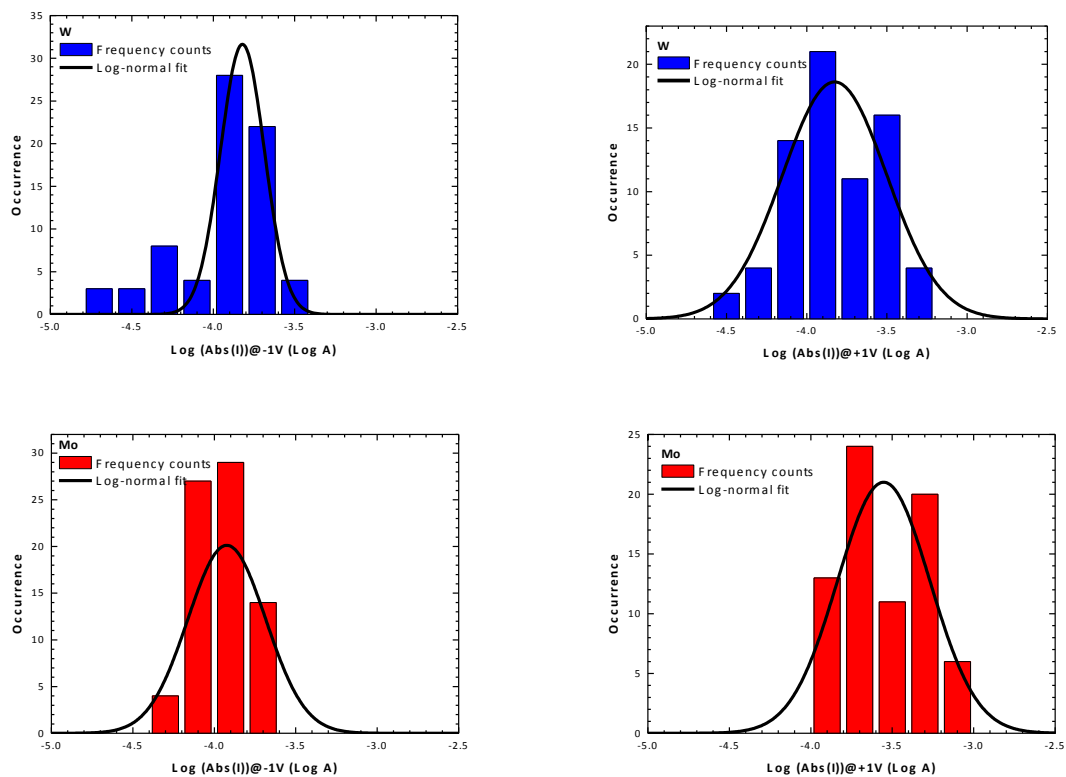

Figure S15. Histograms of the measured current at +1V and -1V for the K$^W_{Sn}$ and K$^{Mo}_{Sn}$ samples. They are fitted by one log-normal distribution. In the case of the K$^W_{Sn}$ monolayer, the mean values of the current histograms at -1V and +1V are -1.52 x 10$^{-4}$ A and +1.50 x 10$^{-4}$ A, respectively (standard deviations of 0.24 and 0.68). For the K$^{Mo}_{Sn}$ monolayer, the mean values of the current at -1V and +1V are -1.09 x 10$^{-4}$ A and +2.72 x 10$^{-4}$ A (with standard variations of 0.34 and 0.64), leading to a mean rectification ratio I$_{+1V}$/I$_{-1V}$ of around 2.5.

## 5. I-V curves adjustment with the Simmon's equation

The expression of the tunnel current through a potential barrier was given by Simmons equation (equation 1):[5]

$$\frac{I}{S} = \frac{e}{4\pi h s^2}\left\{(2\Phi_{POM}-eV)\exp-\frac{4\pi s\sqrt{m(2\Phi_{POM}-eV)}}{h} - (2\Phi_{POM}+eV)\exp-\frac{4\pi s\sqrt{m(2\Phi_{POM}+eV)}}{h}\right\} \quad (eq.\ 1)$$

with $e$ the elementary charge, $h$ Planck's constant, $s$ thickness of the tunneling barrier, $\Phi$ barrier height, $V$ voltage applied to the junction, $m$ effective mass of electron, $I$ current and $S$ the electrical contact surface area. The electron effective mass $m$ is separated in $m = m_r \cdot m_0$ with $m_0$ the mass of electron and $m_r$ the reduced mass.

Adjustments of the measured I-V curves are systematically done for the positive bias (between 0 to 1V, since the I-V curves are quite symmetric with respect of the voltage polarity) by fixing two paramaters (i) $s$ corresponding to the thickness of the monolayer determined by ellipsometry; (ii) and $S$ the surface contact area estimated in our system to ~3 x $10^{-4}$ cm². This adjustment was realized with Origin 2016 from OriginLab Corp (function "Nonlinear Curve Fit" with the "Levenberg Marquardt" Iteration Algorithm).

## 6. Histograms of effective mass obtained by Simmon's adjustments on $K^W_{Sn}$ and $K^{Mo}_{Sn}$ monolayers

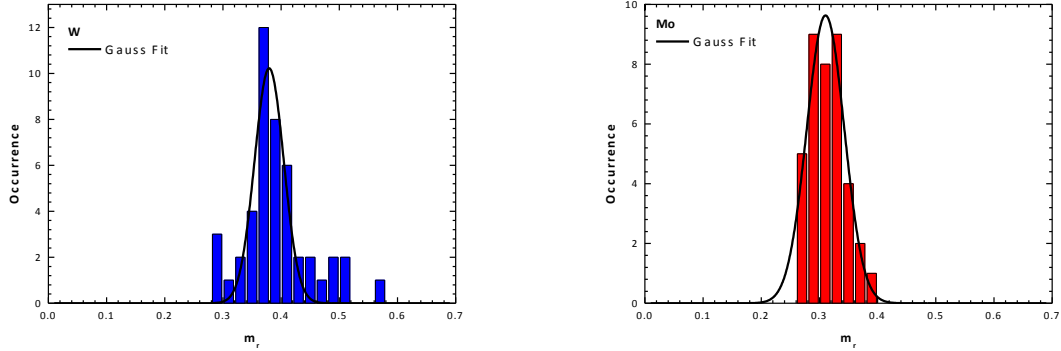

Figure S16. Reduced mass ($m_r$) histograms obtained by Simmon's adjustments on the different I-V curves, and adjusted by a Gaussian distribution for the $K^{Mo}_{Sn}$ and $K^W_{Sn}$ derivatized monolayers. The histograms of $m_r$ are fitted by a Gaussian law. The mean values of $m_r$ for the $K^{Mo}_{Sn}$ and $K^W_{Sn}$ monolayers are 0.38 (standard deviation 0.05) and 0.31 (standard deviation 0.06) respectively.

## 7. AFM image of the neat hydrogenated-Si substrate

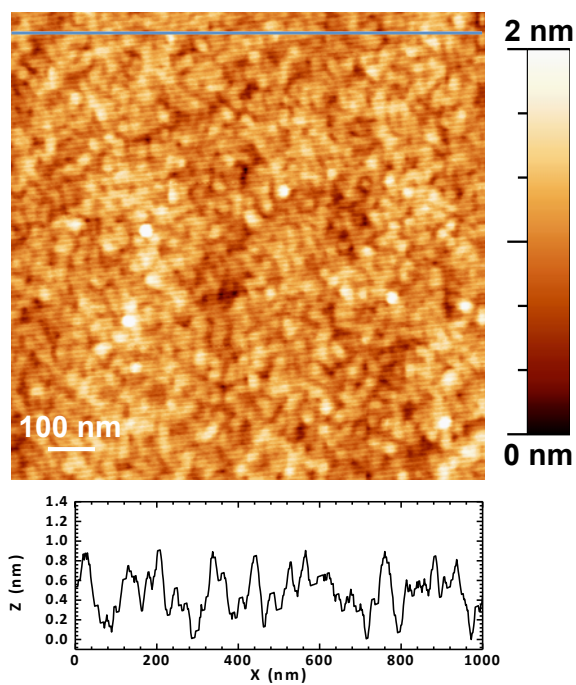

Figure S17. AFM image of the neat hydrogenated-Si substrate (root-mean-squre roughness RMS = 0.25 nm)

**References**
[1] C. Rinfray, G. Izzet, J. Pinson, S. G. Derouich, J. J. Ganem, C. Combellas, F. Kanoufi, A. Proust, *Chem-Eur J* **2013**, *19*, 13838-13846.

[2] C. Rinfray, V. Brasiliense, G. Izzet, F. Volatron, S. Alves, C. Combellas, F. Kanoufi, A. Proust, *Inorg Chem* **2016**, *55*, 6929-6937.

[3] G. Izzet, F. Volatron, A. Proust, *Chem Rec* **2017**, *17*, 250-266.

[4] C. Rinfray, S. Renaudineau, G. Izzet, A. Proust, *Chem Commun* **2014**, *50*, 8575-8577.

[5] J. G. Simmons, *J Appl Phys* **1963**, *34*, 2581-2590.